\documentclass[prl,amsmath,aps,superscriptaddress,twocolumn]{revtex4}
\usepackage{amsfonts}
\usepackage{graphicx}
\usepackage[export]{adjustbox}

\newcommand{\mD}{{\mathcal D}}

\newcommand{\mA}{{\mathcal A}}
\newcommand{\mB}{{\mathcal B}}

\newcommand{\mS}{{\mathcal S}}

\newcommand{\bs}{{\bf s}}

\newcommand{\tr}{{\rm Tr}}
\begin{document}
\title{A family of nonlocal bound entangled states}
\author{Sixia Yu}
\address{Centre for Quantum Technologies, National University of Singapore, 2 Science Drive 3, Singapore 117542}
\address{Hefei National Laboratory for Physical Sciences at
Microscale \& Department of Modern Physics,
 University of Science and Technology of China, Hefei, Anhui 230026, China}
\author{C.H. Oh}
\address{Centre for Quantum Technologies, National University of Singapore, 2 Science Drive 3, Singapore 117542}

\begin{abstract}
Bound entanglement, being entangled yet not distillable, is essential to our understandings of the relations between nonlocality and entanglement besides its applications in  certain quantum information tasks. Recently,  bound entangled states that violate a Bell inequality have been constructed for  a two-qutrit system, disproving a conjecture by Peres that bound entanglement is local.  Here we shall construct such kind of nonlocal bound entangled states for all finite dimensions larger than two, making possible their experimental demonstrations on most general systems. We propose  a Bell inequality, based on a Hardy-type argument for nonlocality, and a steering inequality to identify their nonlocality.  We also provide a family of entanglement witnesses to detect their entanglement  beyond the Bell inequality and the steering inequality.
\end{abstract}
\maketitle

{\it Introduction --- }Quantum nonlocality \cite{bell,review} and entanglement \cite{werner,horo_RMP} are two intricately entwined quantum features that are essential in most quantum information processes in addition to shedding light on our understandings of reality. On the one hand, every entangled pure state is nonlocal \cite{gisin, PR92,dani}, which can be signaled by the violation of a single Bell inequality \cite{yu12}. On the other hand,  we are dealing with mixed states in most cases due to ubiquitous noises and there are entangled mixed states, e.g., Werner's states \cite{werner}, that admit a local realistic model, i.e., cannot violate any Bell inequality. Fortunately, by using distillation protocols  \cite{ED} that involve only local operations and classical communications one can  extract pure entanglement from many copies of entangled mixed states, showing therefore the nonlocality of entangled states that are distillable. 

However there are entangled states, namely bound entangled states \cite{BE}, that are not distillable. This delicate entanglement  does not exist in two-qubit and qubit-qutrit systems and the only examples known so far are entangled states with positive partial transpose (PPT) \cite{PPT, horo96}.  This mystical invention of nature, as called by its founder \cite{horo_RMP}, is  useful in certain quantum communication tasks not achievable by local means, e.g., distilling a secure quantum key \cite{jonathan} and reducing the communication complexity \cite{epping}.
Peres \cite{peres99} conjectured that bound entangled states were local, i.e., cannot violate any Bell inequality, and this conjecture was disproved at first in the multipartite case \cite{Dur,VB} and most recently for a two-qutrit system  by the discovery of a family of bound entangled states \cite{sbe} that violate a Bell inequality \cite{be3}. A stronger version of Peres conjecture \cite{pusey} on the steerability \cite{sch} was also disproved by the same family of states \cite{sbe}.

In this Letter we generalize this  family of nonlocal bound entangled states to all finite dimensions greater than two. We propose a Bell inequality, which comes from a Hardy-type argument, and a steering inequality and identify non-empty sets of nonlocal bound entangled states that give rise to small but finite violations. Our analytical approach also enables us to find the asymptotic violations in the limit of large local dimension. Moreover we present a family of entanglement witnesses to detect their entanglement.

{\it Nonlocality, steerability, and entanglement --- }Let Alice and Bob be two space-like separated observers, each performing some local measurements on the compound system they share. If the correlation $P(a,b|A,B)$ of every pairs of local measurements $A$ and $B$ with outcomes $a,b$ assumes the following local form 
\begin{equation}
P(a,b|A,B)=\sum_{\lambda}P(\lambda)P_{\alpha}(a|A,\lambda)P_\beta(b|B,\lambda)
\end{equation}
with $\alpha,\beta\in\{q,c\}$, then the state of the compound system is separable \cite{werner} in the case of $(\alpha,\beta)=(q,q)$, unsteerable by $A$ or $B$  \cite{steering} if $(\alpha,\beta)=(c,q)$ or $(\alpha,\beta)=(q,c)$, which is also called as a local hidden state model, and local if $(\alpha,\beta)=(c,c)$, which is known as a local hidden variable model. Here for a given hidden variable $\lambda$ distributed according to $P(\lambda)$, we denote $P_q(a|A,\lambda)=\tr(\sigma_\lambda A)$ for some quantum state $\sigma_\lambda$ and a quantum measurement $\{A\}$ and by $P_c(a|A,\lambda)$ a most general probability distribution, including quantum statistics as a special case. If such a local model does not exist, then the state is called as entangled (not $qq$), $A(B)$-steerable (not $qc$ or not $cq$), and Bell nonlocal (not $cc$), respectively. Entanglement is necessary for steerability and steerability is necessary for the nonlocality.
Various kinds of entanglement witnesses \cite{ew}, e.g., via local orthgonal observables \cite{yu05}, steering inequalities \cite{steering}, and Bell inequalities have been proposed to detect the entanglement and nonlocality.

{\it The nonlocal bound entangled states --- }
Consider a bipartite system of two qudits with each qudit having $d\ge 3$ distinguishable states $\{|i\rangle\}_{i=0}^{d-1}$ and denote by $\{|i,j\rangle=|i\rangle\otimes|j\rangle\}_{i,j=0}^{d-1}$ the computational basis of the whole system. Essential to our construction is a set $\Theta_d=\{|\theta_p\rangle\}_{p=0}^{d-1}$ of $d$ normalized  pure states  of a single qudit in the $d-1$ dimensional subspace spanned by $\{|i\rangle\}_{i=1}^{d-1}$  satisfying
\begin{equation}\label{mp}
\langle \theta_p|\theta_q\rangle=-\frac{1}{d-1},\quad (\forall\ p\not=q).
\end{equation}
 \begin{widetext}\mbox{\ }
\begin{figure}
    \vspace{0pt}\includegraphics[scale=1,valign=t]{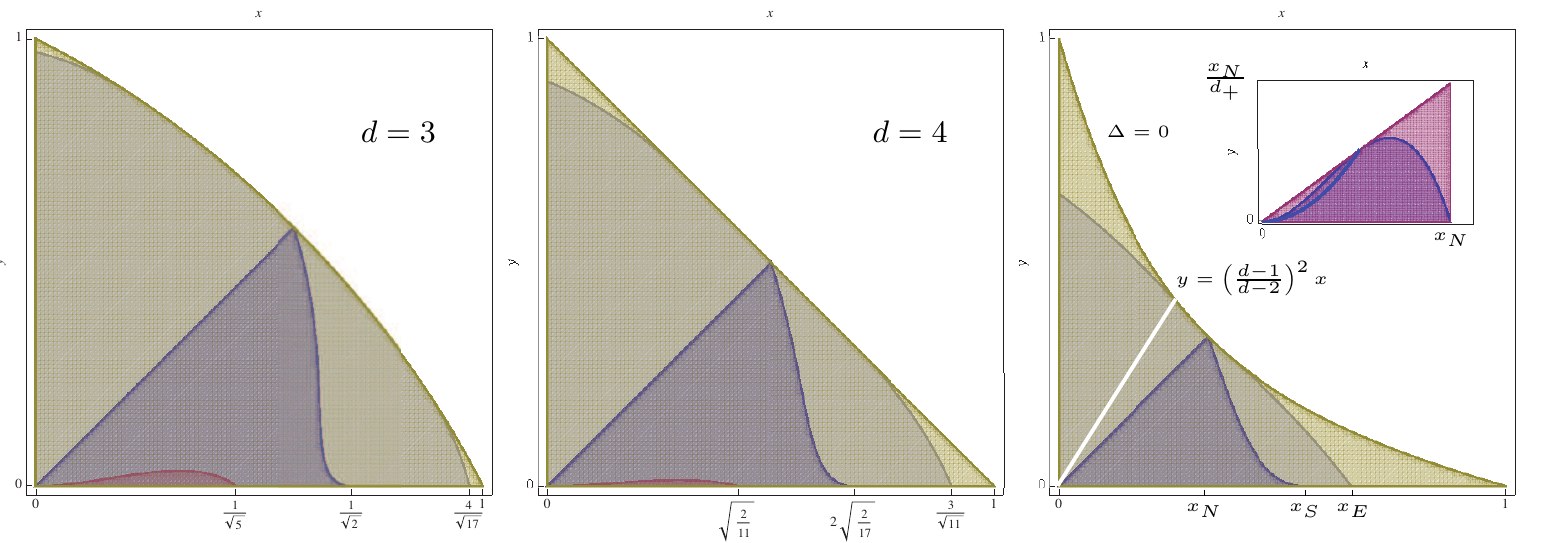}
\caption{(Color online) Bell nonlocal, steerable, bound entangled, and PPT states are illustrated in red, gray, green, and yellow  regions, i.e., $\mD_N^x\subset \mD_S\subset \mD_E\subset \mD$, respectively, in the case of $d=3$ (left) and $d=4$ (center). In the general case of $d\ge 5$ (right) the region $\mD_N^x$ of nonlocal bound states is illustrated in the inset with the blue curve lying inside to show its nonemptiness.}
\end{figure}{}
\end{widetext}
\noindent A recursive construction and basic properties of these highly symmetric states $\Theta_d$ are provided in the supplementary material. Our state reads
\begin{equation}\label{st}
\varrho_{xy}=\frac{xy}{R}|\Psi\rangle\langle\Psi|+\frac\Delta{R}\sum_{{i>j=1}}^{d-1}{|\psi_{ij}\rangle\langle\psi_{ij}|}+\frac1R\sum_{k=1}^{d-1}|\psi_k\rangle\langle\psi_k|
\end{equation}
where $x,y>0$ satisfy $\Delta:={z^2}/({d-2})-xy>0$ with $z=\sqrt{1-x^2-y^2}$ and $R=dxy+(d-1)(d-2)\Delta+d-1$ is the normalization constant, and \begin{subequations}
\begin{eqnarray}
&\displaystyle|\Psi\rangle=\sum_{i=0}^{d-1}|i,i\rangle,\quad|\psi_{ij}\rangle={|i,j\rangle-|j,i\rangle},&
\\
&\displaystyle|\psi_k\rangle=x|0,k\rangle+y|k,0\rangle+z|\phi_k\rangle,&\\
&\displaystyle|\phi_k\rangle=\frac{(d-1)^{\frac32}}{d\sqrt{d-2}}\sum_{p=0}^{d-1}|\theta_p\rangle\otimes|\theta_p\rangle\langle \theta_p|k\rangle.&
\end{eqnarray}
\end{subequations}

We denote by $\mD=\{(x,y)|x,y,\Delta>0\}$ and for each $(x,y)\in \mD$ the state $\varrho_{xy}$ is well defined, with the pure states appearing in its definition as eigenstates, and has positive partial transpose because $\varrho_{xy}^{T_1}=\varrho_{xy}$ as shown in supplementary material.  If $d=3$ our states are equivalent to those nonlocal bound states given in \cite{sbe} under a local unitary transformation $\{|1\rangle\to -|2\rangle,|2\rangle\to|1\rangle\}$ on the first qutrit together with $\{|1\rangle\leftrightarrow|2\rangle|\}$ on the second qutrit, with $|0\rangle$ unchanged.  Our main result reads:

{\bf Theorem } The state $\varrho_{xy}$  is Bell nonlocal if $(x,y)\in \mD_{N}^x\cup \mD_{N}^y$, $B(A)$-steerable if $(x,y)\in \mD_S^x$ $(\mD_S^y)$, respectively, and entangled if $(x,y)\in \mD_E$, where $\mD_{N}^x\subset \mD$ denotes the open set defined by 
\begin{equation}\label{dnl}
\frac{(y\tilde z+y^2-\frac{x^2}{d-1})^2}{x^2+(d-1)y^2}<\frac{(x-d_+ y)(x+d_-y)}{(d-1)^2}
\end{equation}
with $\tilde z=z\sqrt{d-2}$ and $d_\pm=d\sqrt{(d-1)(d-2)}\pm (d-1)^2$ and $\mD_S^x\subset \mD$  denotes the open set defined by conditions i) $x>y>0$ and ii)
\begin{equation}\label{ds}
\frac{(d-1)x+y}2\left(1+\sqrt{\frac yx}\right)<{\tilde z}+2y
\end{equation}
while $\mD_E\subset \mD$ denotes the open set defined by
\begin{equation}\label{de}
\frac{z}{\sqrt{d-2}}>\left\{\begin{array}{ll} \frac{(d-1)^2x+(d-2)^2y}{2(d-1){(d-2)}}& \sqrt {\frac xy}<\frac {d-2}{d-1},\\
\sqrt{xy}& \frac {d-2}{d-1}\le \sqrt {\frac xy}\le \frac {d-1}{d-2},\\
 \frac{(d-1)^2y+(d-2)^2x}{2(d-1)({d-2})}& \sqrt {\frac xy}>\frac {d-1}{d-2}.\end{array}\right.
\end{equation}
The open sets $\mD_{N}^y$ and  $\mD_S^y$ are obtained by exchanging $x,y$ in the definitions of $\mD^x_{N}$ and  $\mD_S^x$ respectively and it holds $\mD_{N}^{x(y)}\subset \mD_S^{x(y)}\subset \mD_E$.

Open sets $\mD_N^{x,y}$ and $\mD_S^{x,y}$ defined above are nonempty for all dimensions since $\mD^x_N$ is nonempty. This is because the curve defined by 
$y\tilde z+y^2=\frac1{d-1}x^2$ with $0<y<x/d_+$, which is shown as the blue curve in the inset of Fig.1, lies inside $\mD^x_{N}$ because the left hand side of Eq.(\ref{dnl}) vanishes  identically  while its right hand side is positive as long as $x\not=d_+ y$. Moreover the open set $\mD^x_{N}$ is contained in the triangle formed by $y=0$, $x=x_N:=\sqrt{(d-2)/(d^2-d-1)}$, and $x=d_+y$ while  $\mD^x_S$ is contained in the triangle formed by $y=0$, $x=y$, and $x=x_S:=2\sqrt{(d-2)/(d^2+2d-7)}$ (see supplementary material). In Fig.1 we have illustrated these open sets, together with $x_N<x_S<x_E:=2(d-1)/\sqrt{d^3-2d^2+4d-4}$ in the case of $d=3,4$ and in the general case of $d\ge 5$.

{\it Bell nonlocality --- }We consider the Bell scenario in which Alice performs $d$ 2-outcome measurements $\mA_p=\{A_p,\bar A_p\}$ with $p=0,1,\ldots,d-1$ while Bob performs one $d$-outcome measurement $\mB=\{B_0,B_1,\ldots, B_{d-1}\}$ and one 2-outcome measurement $\mB^\prime=\{B_0^\prime,B_1^\prime\}$. We shall denote by, e.g., $P(A_pB_q)$ (or $P(\bar A_p B_0^\prime)$) the probability of the event in which Alice measures $\mA_p$ obtaining outcome $0$ (or $1$) and Bob measures $\mB$ (or $\mB^\prime$) obtaining outcome $q$ (or  $0$). In any local realistic model  the following $2d$ conditions cannot be satisfied simultaneously
\begin{subequations}\label{hd}
\begin{eqnarray}
P(A_pB_p)&=&0,\quad (\forall p),\label{hd1}\\
P(\bar A_pB_0^\prime)&=&0,\quad (p\not=0),\label{hd2}\\
 P(A_0B_0^\prime)&>&0.
\end{eqnarray}
\end{subequations}
In fact, any hidden variable triggering the event $A_0 B_0^\prime$, i.e., Alice obtain outcome $0$ when measuring $\mA_0$ and Bob obtains outcome $0$ when measuring $\mB^\prime$, will either cause the measurement $\mA_p$ to have outcome 1 for some $p\not=0$, i.e., conditions Eq.(\ref{hd2}) cannot be satisfied, or cause the measurement $\mA_p$ to have outcome 0 for all $p$, i.e., conditions Eq.(\ref{hd1}) cannot be satisfied since any hidden variable has to trigger one of the event $\{B_p\}$. This Hardy-type of nonlocality test also gives rise to a Bell inequality
\begin{equation}\label{bi}
P(A_0B_0^\prime)-\sum_{p=1}^{d-1}P(\bar A_pB_0^\prime)-\sum_{p=0}^{d-1}P(A_pB_p)\le 0.
\end{equation}
In the case of $d=3$ our Bell inequality is equivalent to the one in \cite{be3} up to some nonsignaling conditions.  Although we fail to detect the nonlocality of our states by using the Hardy-type of argument Eq.(\ref{hd}) we manage to identify a nonempty set of our states that do violate the corresponding Bell inequality Eq.(\ref{bi}).

To this aim we have to properly choose the measurement settings for each party. We consider the following family of basis (which may not be orthogonal)
\begin{equation}
\{|A_p\rangle=a|0\rangle+b|\theta_p\rangle\mid |\theta_p\rangle\in\Theta_d\}
\end{equation}
 for a single qudit with $a,b$ being two arbitrary real numbers satisfying $a^2+b^2=1$. The 2-outcome measurements for Alice are taken to be $\{A_p=|A_p\rangle\langle A_p|,\bar A_p=I-A_p\}$ with $p=0,1,\ldots, d-1$. The orthonormal basis 
\begin{equation}
\{|B_p\rangle=\frac{|0\rangle+\sqrt{{d-1}}|\theta_p\rangle}{\sqrt d}\mid |\theta_p\rangle\in\Theta_d\}
\end{equation}
is taken to be the $d$-outcome measurement for Bob. The 2-outcome measurement $\mB^\prime$ for Bob is simply $\{P_0=|0\rangle\langle 0|,\bar P_0=I-P_0\}$.  Given these measurement settings, we can express the quantum mechanical version of the left hand side of the Bell inequality Eq.(\ref{bi}) as the expectation value of
\begin{equation}
W_{N}=A_0\otimes P_0-\sum_{p=1}^{d-1}\bar A_p\otimes P_0-\sum_{p=0}^{d-1}A_p\otimes B_p
\end{equation}
in the given state $\varrho_{xy}$, which turns out to be, as shown in supplementary material,
\begin{equation}\label{exp}
\tr({\varrho_{xy}} W_{N})=-\frac{d-1}{R}(a,b)M_{N}\left(\begin{array}{c}a\\b\end{array}\right)
\end{equation}
with
\begin{equation}\label{mn}
M_{N}=\left(\begin{array}{cc}x^2+(d-1)y^2&\frac{x(2y+\tilde z)}{\sqrt{d-1}}\\\frac{x(2y+\tilde z)}{\sqrt{d-1}}& \frac{2y\tilde z+\tilde z^2}{d-1}+2xy+(d-2)y^2\end{array}\right).
\end{equation}
In order to violate the Bell inequality Eq.(\ref{bi}) it suffices to demand $\det M_{N}<0$ which turns out to be exactly the condition $(x,y)\in \mD^x_{N}$ determined by Eq.(\ref{dnl}).
By exchanging the roles of Alice and Bob we can obtain a similar Bell inequality from Eq.(\ref{bi}) and similar violations by the state $\varrho_{xy}$ can be obtained if $(x,y)\in \mD^y_N$, since the state $\varrho_{xy}$ is changed into $\varrho_{yx}$ if two qudits are exchanged.

\begin{table}
\begin{tabular}{ccccccccc}
\hline\hline
&$d$ && $(x,y)$ &&$a$&& Max violation&\\\hline
&3&&$(0.309,0.01733)$&&0.913&&2.65264$\times 10^{-4}$\\
&4&&$(0.290,0.00695)$&&0.938&& 7.08492$\times 10^{-5}$\\
&5&&$(0.269,0.00361)$&&0.952&&2.61468$\times 10^{-5}$\\
&6&&$(0.251,0.00218)$&&0.961&& 1.17680$\times 10^{-5}$\\
&7&&$(0.235,0.00141)$&&0.967&&6.05098$\times 10^{-6}$\\
&8&&$(0.222,0.00098)$&&0.971&&3.42082$\times 10^{-6}$\\
&9&&$(0.211,0.00072)$&&0.974&&2.07676$\times 10^{-6}$\\
&$\infty$&&$(\frac 23 d^{-\frac 12},\frac 4{27}d^{-\frac 52})$&&$1-\frac {2}{9d}$&&$ \frac 8{729}d^{-4}$\\
\hline\hline
\end{tabular}
\caption{The maximum violation of Bell inequality by the bound entangled state $\varrho_{xy}$ with measurement settings determined by $a$ in the case of $3\le d\le 9$ and in the large $d$ limit.}
\end{table}

In the cases of $3\le d\le 9$ the maximal violations over all possible nonlocal bound entangled states in $\mD^x_{N}$, together with the optimal $a$ determining the measurements $\{A_p\}$,  are documented in Table I.
The maximization is taken over all the measurements parametrized by some $(a,b)$ as specified  above. Larger violations might be possible by choosing different kind of measurements.  In the case of $d=3$ the analytical counterexample presented in \cite{be3} corresponds to $a=\sqrt{24}/5$ while $x=3/10$ and $y=1/60$. Actually, the violation can be obtain analytically for every single state in $\mD_N^x$ for all dimensions and in the large $d$ limit the maximal violation can also be obtained analytically as shown in supplementary material.

{\it Steerability beyond nonlocality --- } Bell nonlocal states are also steerable. Next we consider the steerability of our states, e.g., the possibility of Bob steering Alice, i.e., $B$-steerablity. For Bob we assume the same measurement settings as in the Bell scenario, i.e.,  $\mB=\{B_p\}_{p=0}^{d-1}$  and  $\mB^\prime=\{B_0^\prime,B_1^\prime\}$. For Alice, since quantum theory is applicable, we consider a  set of $d+3$ positive semidefinite operators $\{Z_{dd},Z_{pd},Z_{d\tau}\}_{p=0}^{d-1}$ satisfying 
\begin{equation}\label{Z}
Z_{dd}-Z_{d\tau}-Z_{pd}\le 0, \quad (\forall\ p,\tau).
\end{equation} 
If the bipartite state is unsteerable from Bob to Alice, it holds the following inequality
\begin{equation}\label{si}
P_A(Z_{dd})-\sum_{\tau=0}^1P(Z_{d\tau} B_\tau^\prime)-\sum_{p=0}^{d-1}P(Z_{pd} B_p)\le 0.
\end{equation}
In \cite{sbe} an additional constraint $Z_{dd}=Z_{0d}$ has been imposed. A slightly larger violation to the above inequality can be expected by a more general choice. We consider the following family of operators
\begin{equation}
\begin{array}{c}Z_{d1}=Z_{dd}=(1-s)a^2P_0,\\
Z_{d0}=(s^{-1}-1)b^2\bar P_0, \quad Z_{pd}=|A_{p}\rangle\langle A_{p}|\end{array}
\end{equation} 
that are parametrized by   $(a,b)$  and $0<s<1$. For any $p,a,b,$ and  $1>s>0$ we have the following inequality, as shown in supplementary material,
\begin{equation}\label{az}
|A_{p}\rangle\langle A_{p}|-(1-s)a^2P_0+(s^{-1}-1)b^2\bar P_{0}\ge 0
\end{equation}
so that the conditions Eq.(\ref{Z}) for $Z$ operators are satisfied. By choosing the same measurement settings for Bob as in the Bell scenario, i.e.,  $\{B_p=|B_p\rangle\langle B_p|\}$ and  $\{B_0^\prime=P_0,B_1^\prime=\bar P_0\}$, the quantum mechanical version of the left hand side of the steering inequality is given by the expectation value of 
\begin{equation}
W_S=Z_{dd}\otimes P_0-Z_{d0}\otimes P_0-\sum_{p=0}^{d-1}Z_{pd}\otimes B_p
\end{equation}
in the given state $\varrho_{xy}$ which assumes the same  form as Eq.(\ref{exp}) with $M_N$ replaced by
\begin{equation}
M_S=\left(\begin{array}{cc}x^2+\frac{sxy}{d-1}&\frac{x(2y+\tilde z)}{\sqrt{d-1}}\\\frac{x(2y+\tilde z)}{\sqrt{d-1}}& \frac{(y+\tilde z)^2}{d-1}+xy+\frac{1-s}sy^2\end{array}\right).
\end{equation}
In order to violate the steering inequality Eq.(\ref{si}) it suffices to demand $\det M_S<0$ for some $0<s<1$. A straightforward calculation yields the conditions $x>y$ and Eq.(\ref{ds}), i.e.,  $(x,y)\in \mD_S^x$ (see supplementary material).  By minimizing the negative eigenvalue of $M_S$  over all possible $s$ with $(a,b)$ taken to be the eigenstate of $M_S$ corresponding to the negative eigenvalue, we obtain the maximal violation for a given state. The maximal violation over all possible states in $\mD_S^x$ for each $3\le d\le 9$ are documented in Table II, as well as the asymptotical maximal violation (see supplementary material).  As expected, in the case of $d=3$ there is a larger violation to the steering inequality Eq.(\ref{si}) than that was found in \cite{sbe} with a restricted measurement setting, which identifies only a subset of steerable states $\varrho_{xy}$ in $\mD^x_S$. 

\begin{table}
\begin{tabular}{ccccccccccccc}
\hline\hline
&$d$ && $(x,y)$ && $s$ && $a$&& Max violation&\\\hline
&3&&$(0.473,0.182)$&&0.5413&&0.851&&3.2655$\times 10^{-3}$\\
&4&&$(0.434,0.154)$&&0.5370&&0.887&&2.0082$\times 10^{-3}$\\
&5&&$(0.400,0.136)$&&0.5370&&0.908&&1.3277$\times 10^{-3}$\\
&6&&$(0.372,0.123)$&&0.5373&&0.923&&9.3813$\times 10^{-4}$\\
&7&&$(0.349,0.114)$&&0.5377&&0.933&&6.9687$\times 10^{-4}$\\
&8&&$(0.330,0.106)$&&0.5380&&0.941&&5.3768$\times 10^{-4}$\\
&9&&$(0.313,0.099)$&&0.5382&&0.947&&4.2729$\times 10^{-4}$\\
&$\infty$&&$(d^{-\frac 12},\frac 14 d^{-\frac 12})$&&$1/2$&&$1-\frac 1{2d}$&&$1/(32d^2)$\\
\hline\hline
\end{tabular}
\caption{The maximal violation of the steering inequality by the PPT state $\varrho_{xy}$ with measurement settings determined by $a$ and $s$ in the case of $3\le d\le 9$.}
\end{table}

{\it Entanglement beyond steerability and nonlocality--- }The violation to the steering inequality as well as the Bell inequality provides naturally an entanglement witness, namely $W_N$ and $W_S$, for the nonlocal bound entangled states $\varrho_{xy}$. These witnesses are however relatively weak with respect to entanglement detection because the quantum nature of none or only one party is taken into account. It turns out that these two witnesses belong to the following family of entanglement witnesses 
\begin{equation}
W_{E}=(1-\alpha) a^2 P_0\otimes P_0-\beta b^2 \bar P_0\otimes P_0-\sum_{p=0}^{d-1} A_p\otimes B_p
\end{equation}
where $\alpha$ and $\beta$ are two real numbers and $P_0,|B_p\rangle$ and $|A_p\rangle$ are defined as before with $a^2+b^2=1$. In fact $W_S$ corresponds to the choice $\alpha_S=s$ and $\beta_S=s^{-1}-1$ with $0<s<1$ while $W_N$ corresponds to the choice $\alpha_N=(d-1)b^2/a^2$ and $\beta_N=(d-1)/b^2-\frac d{d-1}$.

For $W_E$ to be an entanglement witness it should hold $\tr(\rho_{sep}W_E)\le 0$ for all separable states $\rho_{sep}$ or equivalently, $\tr_A [(|\psi\rangle\langle\psi|\otimes I) W_E]\le 0$ for all single qudit pure state $|\psi\rangle$ with the partial trace taken over the first qudit. As shown in supplementary material $W_E$ is an entanglement witness if and only if  $1>\alpha\ge0$ and 
\begin{equation}\label{ab}
\alpha t^2+\beta \ge \gamma(t):=\frac{3t^2-2\frac{d-2}{d-1}t-\frac 1{d-1}}{(d-1)\left(t-\frac{d-2}{d-1}\right)^2+1}
\end{equation}
for $1\le t\le 1+\frac d{(d-1)(d-2)}$. Let $J$ denote the set of all pairs $(\alpha,\beta)$ satisfying the conditions above, as illustrated in Fig.2, and its boundaries are $\alpha=0,1$ and the envelop  
of the straight lines defined by Eq.(\ref{ab}) taking equality
\begin{eqnarray}\label{env}
(\alpha_t,\beta_t)=\left(\frac{\dot\gamma(t)}{2t},\ \gamma(t)-\frac{\dot\gamma(t)}{2}t\right).
\end{eqnarray}

\begin{figure}
    \vspace{0pt}\includegraphics[scale=0.65,valign=t]{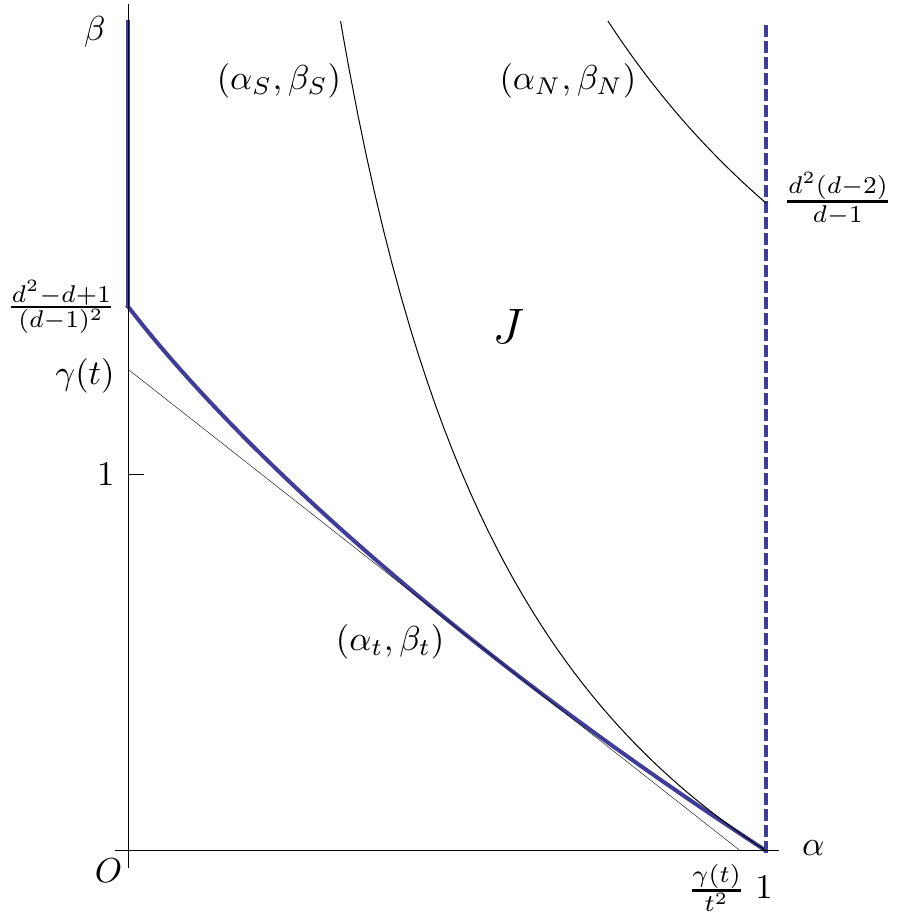}
\caption{(Color online) Illustration (not to the scale) of the range $J$ of $(\alpha,\beta)$ for $W_E$ to be an entanglement witness.}
\end{figure}{}

As expected $(\alpha_S,\beta_S)$ and $(\alpha_N,\beta_N)$ lie in the interior of $J$ and the nontrivial witness on the boundary of $J$, namely $(\alpha_t,\beta_t)$, will detect a larger set of bound entangled states. 
The expectation value of $W_E$ in the state $\varrho_{xy}$ assumes the same form as Eq.(\ref{exp}) with $M_N$ replaced by 
\begin{equation}
M_E=\left(\begin{array}{cc}x^2+\frac{\alpha_txy}{d-1}&\frac{x(2y+\tilde z)}{\sqrt{d-1}}\\\frac{x(2y+\tilde z)}{\sqrt{d-1}}& \frac{(y+\tilde z)^2}{d-1}+xy+\beta_ty^2\end{array}\right).
\end{equation}
Since the state is invariant under the exchanging of two qudits and $x$ and $y$ we can obtain a similar entanglement witness $W^\prime_E$  from $W_E$ by exchanging two qudits. Its expectation value in $\varrho_{xy}$ is determined by the matrix $M_E^\prime$ obtained form $M_E$ by exchanging $x$ and $y$. 
In order to have an entangled PPT state $\varrho_{xy}$ it suffice to have $\det M_E<0$ or $\det M_E^\prime <0$ which turns out to be the condition $(x,y)\in \mD_E$ (see supplementary material).

{\it Conclusions and discussions --- } We have constructed a family of bound entangled states and proposed a Bell inequality, a steering inequality, and a family of entanglement witnesses to detect their nonlocality, steerabilty, and entanglement. Our entanglement witnesses can also help detect other bound entangled states and entangled states for which other criteria might fail. Our proposed bound entangled states may find applications in the nonlocality-based or and semi-device dependent quantum information tasks.  Their  preparation in various physical systems might be facilitated by the symmetry of $\varrho_{xy}$ exhibited via $\Theta_d$.   We believe that all the proposed states are entangled, as suggested by numerical evidences,  even though they cannot be comprehensively detected by our entanglement witness. The questions of its generalization to continuous variable systems and bipartite systems with unequal local dimensions are left open.

{\it Acknowledgement --- } This work is funded by the Singapore Ministry of Education (partly through the Academic Research Fund Tier 3 MOE2012-T3-1-009).

\newpage
\setcounter{equation}{0}
\renewcommand{\theequation}{S\arabic{equation}}
\section{Supplemental Material}
{\it Construction of $\Theta_d$ --- }For examples we have $\Theta_2=\{\pm |1\rangle\}$ and $\Theta_3=\{(\pm \sqrt 3|1\rangle-|2\rangle)/2,|2\rangle \}$. In general $\Theta_d$ with $d\ge 3$ is defined recursively by $\Theta_{d-1}$ via
\begin{eqnarray}
 |\theta_p\rangle_{d-1}=\frac {\sqrt{d(d-2)}|\theta_p\rangle_{d-2}-|d-1\rangle}{d-1}
 \end{eqnarray}
for $0\le p\le d-2$ and $|\theta_{d-1}\rangle_{d-1}=|d-1\rangle$. All the coefficients of $|\theta_p\rangle$ in the computational basis are real numbers, i.e., $\langle \theta_p|k\rangle=\langle k|\theta_p\rangle$ for all $k$. Since the Gramm matrix of those $d$ states in $\Theta_d$ has rank $d-1$ there are exactly $d-1$ independent state in $\Theta_d$ and it holds
 \begin{equation}\label{theta}
\sum_{p=0}^{d-1}|\theta_p\rangle=0,\quad
\sum_{p=0}^{d-1}|\theta_p\rangle\langle \theta_p|=\frac d{d-1}\bar P_{0},
\end{equation} where $\bar P_{0}=I-P_0$ with $P_0=|0\rangle\langle 0|$ is the projection to the $d-1$ dimensional subspace.

{\it Partial transpose of $\varrho_{xy}$ ---} We shall prove that the state $\varrho_{xy}$ defined in Eq.(\ref{st}) has positive partial transpose  for all $(x,y)\in \mD$, i.e., $x,y,\Delta>0$ by showing that $\varrho_{xy}=\varrho_{xy}^{T_1}$ with $T_1$ denoting the partial transpose made on the first qudit. To proceed we introduce a $d-1$ dimensional maximally entangled state
\begin{equation}
|\Phi\rangle:=\sum_{k=1}^{d-1}|k,k\rangle=\frac{d-1}d\sum_{p=0}^{d-1}|\theta_p\rangle\otimes|\theta_p\rangle
\end{equation}
in which we have taken into account Eq.(\ref{theta}). For simplicity we shall denote by a hatted letter, e.g., $\hat\theta_p$, the projection of the corresponding pure state, e.g., $|\theta_p\rangle\langle\theta_p|$, in what follows. First, since $|\Psi\rangle=|00\rangle+|\Phi\rangle$,
we have
\begin{equation}\label{1}
\hat\Psi^{T_1}=|00\rangle\langle 00|+\hat\Phi^{T_1}+\sum_{k=1}^{d-1}(|0,k\rangle\langle k,0|+|k,0\rangle\langle 0,k|).
\end{equation}
Second, from the identity
$$\sum_{i>j=1}^{d-1}\hat\psi_{ij}=\sum_{i,j=1}^{d-1}(|i,j\rangle\langle i,j|-|i,j\rangle\langle j,i|)=\bar P_0\otimes\bar P_0-\hat\Phi^{T_1}$$
it follows  that
\begin{equation}\label{2}
\sum_{i>j=1}^{d-1}(\hat\psi_{ij}-\hat\psi_{ij}^{T_1})=\hat\Phi-\hat\Phi^{T_1}.
\end{equation}
Third, by taking into account the fact that $\sum_{k=1}^{d-1}|k\rangle\langle k|=\bar P_0$, $\bar P_0|\theta_p\rangle=|\theta_p\rangle$, and $\langle\theta_p|\theta_q\rangle=\frac{d\delta_{pq}-1}{d-1}$, we obtain 
\begin{equation}\sum_{k=1}^{d-1}|0,k\rangle\langle\phi_k|=\frac{(d-1)^{\frac 32}}{d\sqrt{d-2}}\sum_{p=0}^{d-1}|0\rangle\langle \theta_p|\otimes |\theta_p\rangle\langle\theta_p|,\end{equation} 
and
\begin{eqnarray}
&&\frac{d^2(d-2)}{(d-1)^{3}}\sum_{k=1}^{d-1}\hat\phi_k\nonumber\\
&=&\sum_{p,q=0}^{d-1} |\theta_p\rangle\langle \theta_q|\otimes |\theta_p\rangle\langle \theta_q|\sum_{k=1}^{d-1} \langle \theta_p|k\rangle\langle k|\theta_q\rangle\nonumber\\
&=&\sum_{p,q=0}^{d-1} |\theta_p\rangle\langle \theta_q|\otimes |\theta_p\rangle\langle \theta_q|\frac{d\delta_{pq}-1}{d-1}\nonumber\\
&=&\frac d{d-1}\sum_{p=0}^{d-1} \hat \theta_p \otimes \hat \theta_p-\frac {d^2}{(d-1)^3}\hat\Phi,
\end{eqnarray}
from which 
it follows that
\begin{eqnarray}\label{3}
\sum_{k=1}^{d-1}(\hat\psi_k-\hat\psi_k^{T_1})=xy\sum_{k=1}^{d-1}|0,k\rangle\langle k,0|+|k,0\rangle\langle 0,k|)\cr
-xy\big(|\Phi\rangle\langle 00|+|00\rangle\langle\Phi|\big)+\frac{z^2(\hat\Phi^{T_1}-\hat\Phi)}{d-2}.
\end{eqnarray}
Putting together Eq.(\ref{1}), Eq.(\ref{2}), and Eq.(\ref{3}) and recalling that $\Delta=\frac{z^2}{d-2}-xy$, we  obtain
$\varrho_{xy}-\varrho_{xy}^{T_1}=0$.

{\it Bounding triangles for $\mD_N^x$ and $\mD_S^x$ ---} If $(x,y)\in \mD^x_N$ then from condition Eq.(\ref{dnl}) it follows that $x>d_+y$ and 
\begin{eqnarray*}
\frac{\tilde z}y&>& \tilde \lambda^2-1-\sqrt{(1+\tilde \lambda^2)(\tilde \lambda-\tilde d_+)(\tilde \lambda+\tilde d_-)}\nonumber\\ &=&\frac{\sqrt{1+\lambda^2}\big(1+\tilde \lambda^2-(\tilde \lambda-\tilde d_+)(\tilde \lambda+\tilde d_-)\big)}{\sqrt{1+\lambda^2}+\sqrt{(\tilde \lambda-\tilde d_+)(\tilde \lambda+\tilde d_-)}}-2\\
&\ge& \tilde\lambda d\sqrt{d-2}+\frac{d(d-1)(d-2)}2-2
>(d-1)\frac xy
\end{eqnarray*}
with $\tilde \lambda=x/(y\sqrt{d-1})$ and $\tilde d_\pm=d_\pm/\sqrt{d-1}=d\sqrt{d-2}\pm (d-1)^{\frac 32}$. As a result we obtain $x< x_N$  and, considering $x>y$, also $(x,y)\in \mD^x_S$ for $(x,y)\in \mD^x_N$ and even if Eq.(\ref{dnl}) is an equality, i.e., $\mD^x_N\subset \mD^x_S$. If $(x,y)\in \mD^x_S$ then  we have condition Eq.(\ref{ds}) which reads
\begin{equation}\label{K}
2+\frac{\tilde z}y>\left(1+\frac 1{\sqrt \lambda}\right)\frac{(d-1)\lambda+1}2:=K_\lambda.
\end{equation}
with $\lambda=x/y$. Because $\lambda>1$ we have $(d-1)\sqrt \lambda+1/\sqrt \lambda> d$ from which it follows 
$2\tilde z> (d-1)x$, i.e., $x<x_S$.  

{\it Derivation of Eq.(\ref{exp}) --- } 
Recalling that $A_p=|A_p\rangle\langle A_p|$ and $\bar A_p=I-A_p$ with $|A_p\rangle=a|0\rangle+b|\theta_p\rangle$  and identity
\begin{equation}\label{aa}
\sum_{p=0}^{d-1}A_p=da^2P_0+\frac {db^2}{d-1}\bar P_0
\end{equation}
we can rewrite 
\begin{eqnarray}
W_{N}&=&A_0\otimes P_0-\sum_{p=1}^{d-1}\bar A_p\otimes P_0-\sum_{p=0}^{d-1}A_p\otimes B_p\nonumber\\
&=&(1-db^2)P_0\otimes P_0-\left(d-1-\frac{db^2}{d-1}\right)\bar P_0\otimes P_0\nonumber\\
&&-\sum_{p=0}^{d-1}A_p\otimes B_p.
\end{eqnarray}
Since $B_p=|B_p\rangle\langle B_p|$ and $|B_p\rangle=a_0|0\rangle+b_0|\theta_p\rangle$ with $a_0=1/\sqrt d$ and $b_0=\sqrt{(d-1)/d}$ and by denoting $|A_p,B_p\rangle=|A_p\rangle\otimes|B_p\rangle$, we have
$\langle \Psi|A_p,B_p\rangle=aa_0+bb_0.$
Since $|A_p,B_p\rangle$ is symmetric and $|\psi_{ij}\rangle$ is antisymmetric in the subspace spanned by $\{|i\rangle\}_{i=1}^{d-1}$  we have $\langle\psi_{ij}|A_p,B_p\rangle=0$ for all $i\not=j$. Furthermore, from the identity
\begin{eqnarray}
\langle\phi_k|A_p,B_p\rangle&=&\frac{(d-1)^{\frac32}}{d\sqrt{d-2}}\sum_{q=0}^{d-1}\langle\theta_q|A_p\rangle\langle\theta_q|B_p\rangle\langle \theta_p|k\rangle\nonumber\\
&=&bb_0\frac{(d-1)^{\frac32}}{d\sqrt{d-2}}\sum_{q=0}^{d-1}\langle\theta_q|\theta_p\rangle^2 \langle\theta_p|k\rangle\nonumber\\
&=&bb_0\langle\theta_p|k\rangle\sqrt{\frac{d-2}{d-1}}
\end{eqnarray}
for each $k=1,2,\ldots, d-1$, where we have used the facts $\langle\theta_q|\theta_p\rangle=\frac{d\delta_{pq}-1}{d-1}$ and $\sum_p|\theta_p\rangle=0$, it follows
\begin{equation*}
\langle\psi_k|A_p,B_p\rangle=\left(x ab_0+y b a_0+z b b_0 \sqrt{\frac{d-2}{d-1}}\right)\langle\theta_p|k\rangle
\end{equation*}
and thus
\begin{equation*}
\sum_{k=1}^{d-1}|\langle\psi_k|A_p,B_p\rangle|^2=\frac{(xa\sqrt{d-1}+yb+\tilde zb)^2}d.
\end{equation*}
As a result  we have
\begin{eqnarray}\label{abw}
\sum_{p=0}^{d-1}\langle A_p\otimes B_p\rangle_{\varrho_{xy}} =\frac{xy}R \left(a+\sqrt{d-1}b\right)^2\cr+\frac1R\left(xa\sqrt{d-1}+(y+\tilde z)b\right)^2\nonumber\\
=\frac{d-1}R(a,b)\left(\begin{array}{cc}x^2+\frac{xy}{d-1}& \frac{x(2y+\tilde z)}{\sqrt{d-1}}\\ \frac{x(2y+\tilde z)}{\sqrt{d-1}}&\frac{(y+\tilde z)^2}{d-1}+xy\end{array}\right)\left(\begin{array}{c}a\\b\end{array}\right).
\end{eqnarray}
Taking into account $a^2+b^2=1$ and 
\begin{equation}\label{p00}
\langle P_0\otimes  P_0\rangle_{\varrho_{xy}}=\frac{xy}R,\quad \langle \bar P_0\otimes  P_0\rangle_{\varrho_{xy}}=\frac{(d-1)y^2}R
\end{equation}
 we finally obtain Eq.(\ref{exp})
with two by two matrix $M_N$ given by Eq.(\ref{mn}).  In the case of steerability $W_S$ and entanglement witness $W_E$, which assume a similar form as $W_N$, we can obtain similar expression of the expectation value $\tr({\varrho_{xy}} W_N)$ as Eq.(\ref{exp}) with $M_N$ replaced by $M_S$ and $M_E$ respectively.

{\it Analytical violation to the Bell inequality ---} To have a nonzero violation we need $\det M_N<0$, which turns out to be exactly $(x,y)\in \mD_N^x$ defined by Eq.(\ref{dnl}), or equivalently,
\begin{equation}
\left|\frac{\tilde z}y+1-\tilde \lambda^2\right|\le \sqrt{(1+\tilde \lambda^2)(\tilde \lambda-\tilde d_+)(\tilde \lambda+\tilde d_-)}:=\Gamma_{\tilde \lambda}
\end{equation}
where $\tilde \lambda=x/(y\sqrt{d-1})$ satisfying $\tilde \lambda>\tilde d_+=d_+/\sqrt{d-1}$. We can parametrize each $(x,y)\in \mD_N^x$ giving rise to a nonlocal bound entangled state with a real number $\tilde \lambda>\tilde d_+$ and an angle $0<\theta<\pi$ as following
\begin{equation}
y=\left(\frac{(\tilde \lambda^2-1-\Gamma_{\tilde \lambda}\cos\theta )^2}{d-2}+(d-1)\tilde \lambda^2+1\right)^{-\frac12}
\end{equation} 
together with $x=y\tilde \lambda \sqrt{d-1}$, by choosing
\begin{equation}
\frac{\tilde z}y+1-\tilde \lambda^2=\Gamma_{\tilde \lambda}\cos\theta
\end{equation}
and recalling that $\tilde z=\sqrt{(d-2)(1-x^2-y^2)}$. The blue curve shown in the inset of Fig.1 corresponds to $\theta=\pi/2$.

Each given $\tilde \lambda>\tilde d_+$ and $\theta\in (0,\pi)$ define a state $\varrho_{xy}$ via $(x,y)$ given above. For this state we have
\begin{equation}
\det M_N=- y^4\Gamma_{\tilde \lambda}^2\sin^2\theta
\end{equation} 
and we take $(a,b)$ to be the eigenstate, which can be analytically determined by $M_N$, corresponding to the negative eigenvalue
\begin{equation}
\frac{2\det M_N}{{\tr M_N}+\sqrt{({\tr M_N})^2-4\det M_N}}
\end{equation} of $M_N$ and we obtain analytically the nonzero violation
 \begin{equation}
\tr({\varrho_{xy}} W_{N})=\frac{2(d-1)R^{-1}y^4\Gamma_{\tilde \lambda}^2\sin^2\theta}{{\tr M_N}+\sqrt{({\tr M_N})^2-4\det M_N}}.
\end{equation}

Though the maximal violations over all possible states in $\mD_N^x$ in the case of finite dimensions can  be carried out only numerically, in the large $d$ limit, we can obtain the asymptotic maximal violation as follows. Since $\tilde d_+\approx 2d^{3/2}$ we choose $\tilde \lambda=(2+\epsilon)d^{3/2}$ for some $\epsilon>0$ then we have $\Gamma_{\tilde \lambda}^2\approx \epsilon(2+\epsilon)^3 d^6$ so that $1/y\approx ((2+\epsilon)^2+\mu\sqrt{\epsilon(2+\epsilon)^3})d^{5/2}$. Because $\tr M_N\approx 1$, and $R\approx 2d$, we obtain the asymptotic violation $\approx \frac 12 y^4\Gamma_{\tilde \lambda}^2\sin^2\theta$ which attains its maximum at $\cos\theta=2/\sqrt 5$ and $\epsilon=5/2$ giving rise to the asymptotic maximal violation $\approx \frac 8{729}d^{-4}$. The optimal measurement setting reads $a\approx 1-x^2/2$, which is determined by the corresponding eigenstate.

{\it Proof of Eq.(\ref{az}) --- } The inequality holds outside the 2-dimensional subspace spanned by $\{|0\rangle,|\theta_p\rangle\}$ and within this subspace the left hand side of Eq.(\ref{az}) becomes
\begin{equation}
\left(\begin{array}{cc}s a^2& ab\\ ab& b^2/s\end{array}\right)\ge 0.
\end{equation}

{\it Derivation of Eq.(\ref{ds}) --- }  From $\det M_S<0$ for some $0<s<1$ it follows that\begin{eqnarray}
\frac{x^2(2y+\tilde z)^2}{d-1}&>& x^2 L+\frac{xy^3}{d-1}+s \frac{x y L}{d-1}+\frac 1s x^2y^2\nonumber \\
&\ge &(x\sqrt L+y\sqrt{xy/(d-1)})^2
\end{eqnarray}
where we have denoted
$L=\frac{(y+\tilde z)^2}{d-1}+xy-y^2$ and taken $s=\sqrt{(d-1)xy/L}$ to equalize the second inequality. By denoting $\lambda=x/y$ and $v=\tilde z/y$ we obtain
\begin{equation}
{2+v}>\sqrt {(1+v)^2+(d-1)(\lambda-1)}+\frac1{\sqrt\lambda}
\end{equation}
from which it follows
\begin{equation}\label{s1}
1-\frac1{\sqrt{\lambda}}>\frac{(d-1)(\lambda-1)}{\sqrt {(1+v)^2+(d-1)(\lambda-1)}+1+v}.
\end{equation}
If $\lambda\le 1$ then from Eq.(\ref{s1}) it follows 
$1+v\le (d-1)(\lambda+\sqrt\lambda)$ and $2+v<K_\lambda$. Taking into account $v\ge (d-2)\sqrt\lambda$ we obtain a contradiction
\begin{equation}
\frac1{d-1}\le\lambda+\frac{\sqrt\lambda}{d-1}\le\lambda+\frac{2\sqrt\lambda}{d-1}<\frac1{d-1}
\end{equation}
so that we have $\lambda>1$. If $1+v\le (d-1)(\lambda+\sqrt\lambda)$ then  Eq.(\ref{ds}) follows from Eq.(\ref{s1}). If $1+v>(d-1)(\lambda+\sqrt\lambda)$ then Eq.(\ref{ds}) follows from $K_\lambda\le (d-1)\lambda+1<1+v$ since $\lambda>1$. Thus from $\det M_S<0$ for some $0<s<1$ it follows condition Eq.(\ref{ds}) and $x>y$ and vice versa. 

{\it Asymptotic violation to the steering inequality --- } Let us denote
$\lambda=x/y$ and from the condition Eq.(\ref{ds}) for steerability it follows that  there is $\nu>1$ such that
\begin{equation}
y=\frac{\sqrt{d-2}}{\sqrt{(\nu K_\lambda-2)^2+(d-2)(1+\lambda^2)}}
\end{equation}
with $K_\lambda$ defined in Eq.(\ref{K}).
That is to say every pair $(x,y)\in\mD^x_S$ is characterized by two real numbers $\lambda,\nu>1$. In the large $d$ limit we have $1/y\approx \frac12\nu (\lambda+\sqrt \lambda)\sqrt d$ and therefore the largest eigenvalue of $-M_S$ approaches
\begin{eqnarray}
-\det M_S&\approx& xy\left(\frac{2x}{\sqrt d}-x^2+xy-\frac s{d}-\frac{xy}s\right)
\nonumber\\
&\le& \frac{16(\nu-1)(\sqrt \lambda-1)}{\nu^4(\sqrt \lambda+1)^3d^2}
\end{eqnarray}
attains its maximum at $\nu=4/3$ and $\lambda=4$, yielding the asymptotic violation  as listed in Table II. We have the optimal $s=\sqrt{dxy}$ to attain the above inequality.

{\it Entanglement witness --- } For $W_E$ to be candidate of entanglement witness it should hold for every  pure state $|\psi\rangle$ of the first qudit that  $\tr_A [(\hat\psi\otimes I) W_E]\le 0$, i.e.,
\begin{equation}\label{cd1}
\sum_{p=0}^{d-1} s_p B_p\ge h_{\alpha\beta}P_0
\end{equation}
with 
$h_{\alpha\beta}=(1-\alpha)a^2s-\beta b^2\bar s,$
where $s_p=|\langle \psi|A_p\rangle|^2$ and $s=|\langle \psi|0\rangle|^2$ with $\bar s=1-s$. If $s=1$, i.e., $|\psi\rangle=|0\rangle$, then $s_p=a^2$ so that we obtain the condition $\alpha\ge 0$. In the case of $s\not=1$ we introduce
\begin{equation}t:=\frac{|a|\sqrt s}{|b|\sqrt{\bar s}}.\end{equation}
If $t\le 1$ then for any given $p$ we can always choose $|\psi\rangle$ such that $s_p=0$. Thus we have only to require $(1-\alpha)t^2\le \beta$ for all $t\le 1$, from which it follows $\beta\ge 0$, to ensure Eq.(\ref{cd1}) in this case. As a result we obtain the condition $\alpha<1$ otherwise $W_E$ would be negative semi-definite. Now we consider $t> 1$ and in this case 
$${|\langle \psi|A_p\rangle|}{}=|a\langle \psi|0\rangle+b\langle\psi|\theta_p\rangle|\ge |a|\sqrt s-|b|\sqrt{\bar s}>0,$$
since $|\langle\psi|\theta_p\rangle|\le \sqrt {\bar s}$, so that we always have $s_p>0$ for all $p$, i.e., 
$B_\psi=\sum_p{s_pB_p}$ is of full rank. The condition Eq.(\ref{cd1}) now becomes equivalent to 
\begin{eqnarray}\label{ub1}
1\ge h_{\alpha\beta}\langle 0|B_\psi^{-1}|0\rangle= \frac{(1-\alpha) t^2-\beta}d \sum_{p=0}^{d-1}\frac1{\tilde s_p}
\end{eqnarray}
where we have denoted
$\tilde s_p={s_p}/({b^2\bar s})$
for which it holds
\begin{equation}\label{cst1}
\tilde s_p\ge (t-1)^2,\quad \sum_{p=0}^{d-1} \tilde s_p=dt^2+\frac d{d-1}.
\end{equation}

We denote by $\mS_t$ the simplex of $\tilde \bs=(\tilde s_0,\tilde s_1,\ldots,\tilde s_{d-1})$ defined by two conditions above for a given $t$. The extremal points of $\mS_t$ are of form
$$ [\tilde \bs]_p=(t-1)^2, \quad [\tilde \bs]_q=\left(t+\frac 1{d-1}\right)^2:=t_+\quad (q\not= p)$$
for each $p=0,1,\ldots,d-1$. This is because firstly for the extremal point there is at least one $p$ such that $\tilde s_p=(t-1)^2$, due to condition Eq.(\ref{cst1}), and in this case it holds $a\langle\psi|0\rangle=-tb\langle\psi|\theta_p\rangle$ and $|\langle\psi|\theta_p\rangle|^2=\bar s$. Secondly, for a general $\tilde \bs=(\tilde s_0,\tilde s_1,\ldots,\tilde s_{d-1})\in\mS_t$ with a component, say $\tilde s_p$, being equal to $(t-1)^2$, it holds the inequality
$$t_+=\sum_{q\not=p}\frac{\tilde s_q}{d-1}\ge\left(\frac{\sum_{q\not=p}\sqrt{\tilde s_q}}{d-1}\right)^2\ge\frac{\left|\sum_{q\not=p}\langle\psi|A_q\rangle\right|^2}{(d-1)^2b^2\bar s}=t_+$$
where the first equality stems form Eq.(\ref{cst1}), from which it follows $\tilde s_q=t_+$ with $q\not=p$. As a result we obtain
\begin{equation}\label{ub} 
\max_{\tilde\bs\in\mS_t}\sum_{p=0}^{d-1}\frac1{\tilde s_p}=\frac1{(t-1)^2}+\frac{d-1}{\left(t+\frac1{d-1}\right)^2}:=\tilde\gamma(t)
\end{equation}
because the left hand side is a convex function of $\tilde \bs$ so that its maximum in the simplex $\mS_t$ is attained at the extremal points of $\mS_t$.
Since $\gamma(t)=t^2-d/\tilde \gamma(t)$ we obtain condition Eq.(\ref{ab}) from Eq.(\ref{ub1}) together with Eq.(\ref{ub}).  
 
 That is to say $W_E$ is a possible entanglement witness, i.e., nonpositive on all separable states,  if and only if the condition Eq.(\ref{ab}) holds for all $1\le t\le 1+\frac d{(d-1)(d-2)}:=t_1$ with $t_1$ being the unique solution to $\alpha_t=0$ with $t\ge1$ where, with $u=(d-1)(t-1)$,
\begin{eqnarray}
\alpha_t&=&\frac{(d-1)(d+2u)(d-(d-2)u)}{(d+u-1)(d+2u+u^2)^2},\\
\beta_t&=&\frac{u(d+u)(d^2+3du+3u^2)}{(d-1)(d+2u+u^2)^2},
\end{eqnarray}
is the envelop, given by Eq.(\ref{env}), of the straight lines in the $(\alpha,\beta)$ plane defined by Eq.(\ref{ab}) taking equality. It is straightforward to check that $W_E$ with $(\alpha,\beta)\in J$ can detect the entanglement of $|\Psi\rangle$.

\begin{figure}
    \vspace{0pt}\includegraphics[scale=0.5,valign=t]{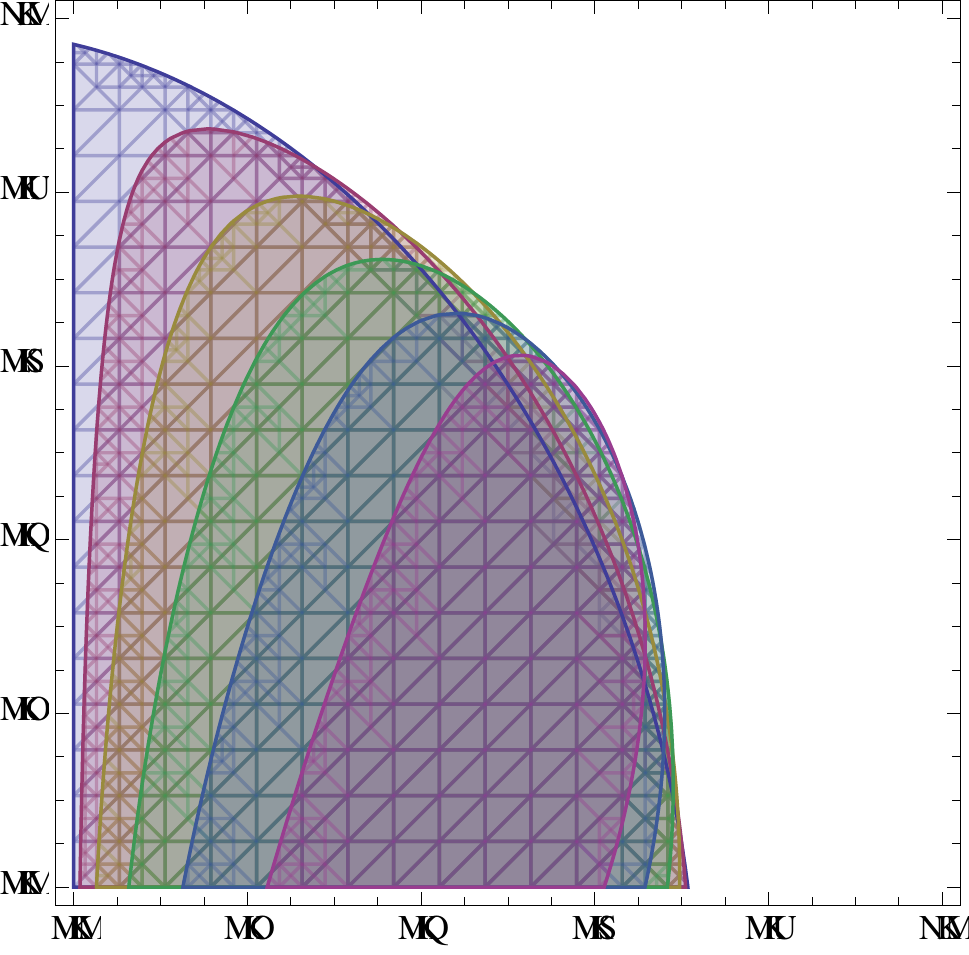}
\caption{(Color online) Illustration of the entanglement region $\mD_E^x$ in the case of $d=3$ that is defined by the envelop of $\det M_E<0$ with $1\le t\le t_1$.}
\end{figure}

By applying the entanglement witness $W_E$ on the boundary of $J$, i.e., $(\alpha_t,\beta_t)$ with $1\le t\le t_1$, to the state $\varrho_{xy}$ we obtain a similar expression of the expectation value $\tr({\varrho_{xy}} W_E)$ as Eq.(\ref{exp}) with $M_N$ replaced by $M_E$. Thus the state is entangled if $\det M_E<0$ for some $1\le t\le t_1$ so that the region enclosed by the envelop of the family of curves in the $x,y$ plane defined by $\det M_E=0$ parametrized by $t$, which is illustrated in Fig.3 in the case of $d=3$, gives rise to bound entangled states. From the equation $\det M_E=0$, i.e.,  
$$\frac{x^2(2y+\tilde z)^2}{d-1}=\left(x^2+ \frac{\alpha_txy}{d-1}\right)\left(\frac{(y+\tilde z)^2}{d-1}+xy+\beta_t y^2\right),$$
and its derivative  with respect to $t$ the envelop is determined by the following two equations
\begin{eqnarray}
(d-1)\lambda+{\alpha_t}=\frac{(2+v)}{t}\sqrt{\lambda},\label{u}\\
\frac{(1+v)^2}{d-1}+\lambda+\beta_t =(2+v)t\sqrt{\lambda},\label{v}
\end{eqnarray}
with $\lambda=x/y$ and $v=\tilde z/y$. From
 Eq.(\ref{u}) it follows that
$v=(d-1)t\sqrt\lambda+{\alpha_tt}/{\sqrt\lambda}-2$
so that Eq.(\ref{v}) becomes a quartic equation $(r-r_0)f(r)=0$ of $r=\sqrt \lambda$ where 
\begin{equation}
r_0=\frac{d+2u}{d+2u+u^2},\quad u=(d-1)(t-1)
\end{equation} and 
\begin{equation}
f(r)=r^3-r^2-\frac{d+2u-du}{(d+2u+u^2)^2}g(r/r_1)
\end{equation}
in which
\begin{eqnarray}
g(r)=1-{(1+u)r}+\left((1+u)^2-\frac1{r_0}\right)r^2\nonumber\\
=\left((1+u)^2-\frac1{r_0}\right)(r-r_c)^2+g(r_c)
\end{eqnarray}
is a quadratic function of $r$ whose minimum 
\begin{equation}
g(r_c)=\frac{\frac34(1+u)^2-\frac1{r_0}}{(1+u)^2-\frac1{r_0}}
\end{equation}
is attained at $r=r_c$ where
\begin{equation}
r_1=\frac{d+2u-du}{d+2u+u^2},\quad r_c=\frac12\frac{1+u}{(1+u)^2-\frac 1{r_0}}.
\end{equation}

As will be shown below $f(r)<0$ for $r<1$ so that we obtain the unique solution to Eq.(\ref{u}) and Eq.(\ref{v}) for $r<1$ as $r=r_0$, i.e.,
\begin{equation}\label{solu}
\sqrt{\frac x y}=r_0,\quad v= \frac{z\sqrt{d-2}}y={(d-2)}\sqrt{\frac xy}
\end{equation}
with $0\le u\le d/(d-2)$, which is exaclty the curve $\Delta=0$ with $(d-2)/(d-1)\le \sqrt{x/y}\le 1$. Together with the curve defined by $\det M_E=0$ with $t=1+d/(d-1)(d-2)$ the envelop Eq.(\ref{solu}) gives rise to Eq.(\ref{de}) in the case of $x<y$. If we consider $W_E^\prime$ with two qudits exchanged we obtain in the same manner Eq.(\ref{de}) in the case of $x>y$.

Now we shall prove $f(r)<0$ when $r<1$. It suffices to show that $g(r/r_1)\ge 0$ for $r<1$ which is true if 
$g(r_c)\ge 0$. If $g(r_c)<0$ then, since $1/{r_0}<1+u,$ it holds $\frac 34(1+u)^2<1/r_0<1+u$ from which it follows $u<1/3$ and
\begin{eqnarray}
2(d+2u-du)-(1+u)(d+2u+u^2)\nonumber\\
=(1-3u)d+2u-3u^2-u^3>0
\end{eqnarray}
i.e., $r_1>({1+u})/2$. As a result we obtain $r_c>2/(1+u)>1/r_1$, considering $\frac 34(1+u)^2< 1/r_0$, so that the function $g(r/r_1)$ of $r$ is decreasing for $r\le 1$. Thus
\begin{eqnarray*}
g(r/r_1)\ge g(1/r_1)=1-\frac{1+u}{r_1}+\frac{(1+u)^2-1/r_0}{r_1^2}\nonumber\\
\ge \frac 1{r_1}\left(r_1+u-\frac 1{r_0}\right)\ge 0.
\end{eqnarray*}

\newpage

\end{document}